\documentclass[iop,twocolappendix,numberedappendix,appendixfloats,tighten,apj]{emulateapj}
\usepackage{apjfonts}

\begin{document}

\title{Enhanced Magnetic Compressibility and Isotropic Scale-Invariance\\* at Sub-Ion Larmor Scales in Solar Wind Turbulence}

\author{K. H. Kiyani\altaffilmark{1,2}, S. C. Chapman\altaffilmark{2}, F. Sahraoui\altaffilmark{3}, B. Hnat\altaffilmark{2}, O. Fauvarque\altaffilmark{1,4} and Yu. V. Khotyaintsev\altaffilmark{5}}

\email{k.kiyani@imperial.ac.uk}

\affil{$^1$Dept. of Electrical and Electronic Engineering; Imperial College London, London, SW7 2AZ, United Kingdom}

\affil{$^2$Centre for Fusion, Space and Astrophysics; University of Warwick, Coventry, CV4 7AL, United Kingdom}

\affil{$^3$Laboratoire de Physique des Plasmas, Observatoire de Saint-Maur, Saint-Maur-Des-Foss{\'e}s, 94107 France}

\affil{$^4$D\'epartement de Physique, Ecole Normale Sup\'erieure, 24 rue Lhomond, 75005 Paris, France}

\affil{$^5$Swedish Institute of Space Physics, Uppsala, Sweden}

\begin{abstract}
	The anisotropic nature of solar wind magnetic turbulence fluctuations is investigated scale-by-scale using high cadence \emph{in-situ} magnetic field measurements from the Cluster and ACE spacecraft missions. The data span five decades in scales from the inertial range to the electron Larmor radius. In contrast to the inertial range, there is a successive increase towards isotropy between parallel and transverse power at scales below the ion Larmor radius, with isotropy being achieved at the electron Larmor radius. In the context of wave-mediated theories of turbulence, we show that this enhancement in magnetic fluctuations parallel to the local mean background field is qualitatively consistent with the magnetic compressibility signature of kinetic Alfv{\'e}n wave solutions of the linearized Vlasov equation. More generally, we discuss how these results may arise naturally due to the prominent role of the Hall term at sub-ion Larmor  scales. Furthermore, computing higher-order statistics, we show that the full statistical signature of the fluctuations at scales below the ion Larmor radius is that of a single isotropic globally scale-invariant process distinct from the anisotropic statistics of the inertial range. 
\end{abstract}

\keywords{plasmas --- magnetic fields --- solar wind --- turbulence --- data analysis}

\maketitle


\section{Introduction} \label{sec:Intro} \emph{In-situ} measurements of fields and particles in the interplanetary solar wind provide unique observations for the study of a turbulent plasma \citep{TuMarsch1995,Horbury2005,bruno2005}. As in neutral fluid turbulence, an inertial range of magnetohydrodynamic (MHD) turbulence is indicated by the observations of a power spectral density (PSD) exhibiting a power-law form over a large range of scales. This power-law is a manifestation of  a scale-invariant turbulent cascade of energy from large to small scales. Unlike neutral fluids where dissipation is carried out by viscosity, interplanetary space plasmas are virtually collisionless. On spatial scales characteristic of the ions there is a transition in the PSD from the inertial range at lower frequencies to a steeper power-law spanning up to two decades in scale to the electron scales \citep{Sahraoui2009, kiyani2009b}. This second interval of scaling in the PSD was dubbed the \emph{dissipation range} in analogy with hydrodynamic turbulence \citep{Leamon1999}. The nature of the fluctuations on these kinetic scales, the mechanisms by which the turbulent energy is cascaded and dissipated, and the possible role of dispersive linear wave modes \citep{Leamon1999,Bale2005,Gary2008,Gary2010} are all hotly debated as vital ingredients of any future model of this dissipation range \citep{Schekochihin2009,Alexandrova2008c,Howes2008b,Sahraoui2010}. 

Anisotropy, with respect to the background magnetic field is a central feature of plasma turbulence in the solar wind \citep{Matthaeus1996,Osman2007,Narita2010,Sahraoui2010}, with the fluctuating components transverse and parallel to the background magnetic field displaying manifest differences in both their dynamics and statistics. The seminal study of \citet{BelcherDavis1971} used Mariner 5 observations to investigate the field variance tensor projected onto an orthonormal `field-velocity' coordinate system. They found that the majority of the power is in the transverse fluctuations with a ratio between transverse and parallel (compressible) components $\sim9:1$. As first seen in high cadence WIND spacecraft data \citep{Leamon1998b}, this ratio decreases to $\sim5:1$ in the dissipation range, indicating enhanced magnetic compressible fluctuations in this sub-ion Larmor scale range. This result showing an enhanced level of magnetic compressibility was repeated with a large set of ACE spacecraft data intervals \citep{Hamilton2008} and also in much higher cadence data from the Cluster spacecraft \citep{Alexandrova2008c}.

In this article we conduct a \emph{scale-by-scale} study of the anisotropy in transverse and parallel fluctuations over five decades in temporal scales from a few hours to tens of milliseconds. Our background magnetic field is scale-dependent, computed self-consistently with the scale-dependent fluctuations as in \citep{Horbury2008,Podesta2009} i.e. a local -- as opposed to global -- background magnetic field. We show that the findings of \citep{Leamon1998b,Alexandrova2008c} are in fact the result of a successive \emph{scale-invariant} reduction in the power ratio between the two components as we move to smaller scales in the dissipation range. This reduction results in isotropy between all three components of magnetic field fluctuations (two transverse and one parallel) at the electron Larmor radius $\rho_{e}$ i.e. equipartition of magnetic energy between all three magnetic field components. Using a Hall MHD model we provide a simple description of how the Hall term in the generalised Ohm's law is responsible for the rise of parallel (compressible) magnetic fluctuations and the resultant isotropy at scales around $\rho_{e}$. In addition, by comparing with linear kinetic solutions of the Vlasov equation and computing the magnetic compressibility \citep{Gary2009a} we show that this scale-invariant reduction in the power ratio is also qualitatively consistent with the transition from shear or highly oblique Alfv\'enic fluctuations in the inertial range possessing no magnetic compressibility, to kinetic Alfv\'en wave-like fluctuations which rapidly develop magnetic compressible components at scales around the ion Larmor radius \citep{Gary2009a,Sahraoui2012}. This nature of the magnetic compressibility is also suggested in the recent work by \citet{Salem2012} who, using a global background field, also computed the magnetic compressibility and compared it with wave solutions of the linearised warm two-fluid equations. Importantly, by calculating higher order statistics and the probability density functions (PDF) we show for the first time that the dissipation range magnetic fluctuations are characterised statistically by a single \emph{isotropic} signature; in stark contrast to the anisotropic inertial range.

We stress that the anisotropy discussed in this article is in the full vector direction of field fluctuations, $\delta\mathbf{B}$, projected parallel and transverse to the local magnetic field direction (also known as \emph{variance anisotropy} \citep{BelcherDavis1971}). This is in contrast to the anisotropy measured in the vector $\mathbf{r}$ which characterises the spatial scale of the fluctuations (or wavevector $\mathbf{k}$ in Fourier space). In the context of the latter anisotropy, the spatial variations in the magnetic field that we measure are a sum of all the components of $\mathbf{k}$ which are projected along the spacecraft trajectory through the solar wind i.e. along the solar wind velocity unit vector $\mathbf{-\hat{V}_{sw}}$; these are then Doppler shifted (via the Taylor frozen-in-flow hypothesis) to spacecraft frame frequencies, $f$, resulting in a \emph{reduced spectrum} 
$\tilde{P}(f,\theta_{BV})=\int d^{3}\bold{k} P(\mathbf{k})\delta(2 \pi f -\mathbf{{k}}\cdot\mathbf{{V}_{sw}}),$
where the $\delta$-function is doing the `reducing', and $\theta_{BV}$ shows the explicit dependence of the reduced spectra on the angle between the background magnetic field and the bulk solar wind velocity vectors (see \citet{Forman2011} and  \citet{Fredricks1976} for further details). The full $\mathbf{k}$ spectrum, $P(\bold{k})$, can only be measured by multi-spacecraft techniques which take into account the full 3D spatial variation of the magnetic field e.g. \emph{k}-filtering \citep{Pincon1991}. In the context of the reduced spectrum, the magnetic field observations investigated in this article are dominated by wavevector components which are strongly oblique ($70^{\circ}$ - $90^{\circ}$) to the background magnetic field; as indicated by the angle between the background magnetic field and solar wind velocity vectors. In such intervals, the inertial range is ubiquitously seen to have a spectral index of $\sim-5/3$, whilst in the dissipation range this steepens to $\sim-2.8$ \citep{kiyani2009b,Alexandrova2009,Chen2010b,Sahraoui2010}. Also, for such intervals, multispacecraft observations of the full $P(\bold{k})$ \citep{Narita2010,Sahraoui2010} suggest that most of the power is in the $k_{\perp}$ rather than the $k_{\parallel}$ components, suggesting that highly oblique angled components of $\mathbf{k}$ are very much representative of most of the magnetic field fluctuation power.

A brief synopsis of the paper is as follows: Section \ref{sec:Data} describes the spacecraft data and summarises the plasma parameters involved, as well as the analysis methods used; Section \ref{sec:Results} forms the bulk of the paper and is split between a.) a study and discussion of the rising magnetic compressibility and component anisotropy in the dissipation range and b.) an anisotropic study of the higher-order statistics. Finally Section \ref{sec:Conclusions} summarises our findings and brings together the insights obtained from these results, concluding with an outlook for future investigations.  Details of the Undecimated Discrete Wavelet Transform that we use to decompose the magnetic field into scale-dependent background field and fluctuations, for use in this study, are described in Appendix \ref{sec:UDWT}.


\section{Observations and Methods} \label{sec:Data} We discuss an interval of quiet ambient solar wind for which there are observations from both the Cluster and ACE spacecraft missions \citep{Escoubet1997,ACEScienceCenter} (all in GSE coordinates). The Cluster (spacecraft 4) interval at 450 Hz cadence (same interval as in \citep{kiyani2009b}) is of an hour duration 2007/01/30 00:10-01:10 UT when the instruments were operating at burst mode, and will primarily be used to study the dissipation range at spacecraft frequencies above 1 Hz. We construct a combined data set from the DC magnetic field (sampled at 67 Hz) of the Flux Gate Magnetometer (FGM) for frequencies below 1 Hz, with the high frequency (oversampled at 450 Hz) search-coil magnetometer data from the STAFF-SCM experiment for frequencies above 1 Hz, using the wavelet reconstruction method \citep{Alexandrova2004,Chen2010b}. The ACE magnetometer (MAG) interval at 1 Hz cadence is over two days 2007/01/30 00:00 UT to 2007/01/31 23:59 UT and samples the inertial range at spacecraft frequencies below 1 Hz. The ACE interval was only needed in order to obtain better estimates of the higher-order statistics in the inertial range (as the Cluster FGM interval is relatively short for this), and thus will only be restricted to this study in the penultimate section of the paper; for the other studies of the inertial range, the FGM data interval was used. 

Both these intervals are in stationary fast wind ($\simeq667\ km\ s^{-1}$) with similar plasma parameters and are sufficiently large data sets so that the sample size variance errors in the computed statistics are negligible \citep{kiyani2009a}. The following plasma quantities are from the Cluster FGM, CIS, PEACE and WHISPER instruments: average magnetic field $\overline{B}\simeq4.5\ nT$, electron plasma $\beta_{e}\simeq1.2$, plasma density $n_{e}\simeq4\ cm^{-3}$, Alfv\'en speed $V_{A}\simeq50\ km\ s^{-1}$, perpendicular ion temperature $T_{i\perp}\simeq24\ eV$, electron temperature $T_{e}\simeq22\ eV$, and ion and electron Larmor radii $\rho_{i}\simeq110\ km$ and $\rho_{e}\simeq1.7\ km$ respectively. The interval is free from any ion or electron foreshock effects at the Earth's bowshock. Moreover, using high-resolution proton moments computed from the 3DP instrument \citep{Lin1995} onboard the Wind spacecraft at 1AU and at the same interval as Cluster, the proton plasma parameters (parallel and perpendicular proton temperatures $T_{p\parallel}\simeq26\ eV$ and $T_{p\perp}\simeq34\ eV$, parallel proton plasma $\beta_{p\parallel}\simeq1.2$) indicate that the intervals are stable to proton pressure anisotropy-driven instabilities \citep{Bale2009}. From the ACE SWEPAM instrument the ion plasma beta $\beta_{i}\simeq1.5$ showing that the ion-inertial length $\lambda_{{i}}\simeq\rho_{{i}}$.

There is a growing opinion \citep{Chapman2007, Horbury2008, Podesta2009, Luo2010} that it is a local scale-dependent mean field consistent with the scale dependent fluctuations, rather than a large scale global field, which should be used in studies of anisotropic plasma turbulence. This is not only self-consistent but also assures that there are no significantly large spectral gaps between the frequencies of the fluctuations being studied and the mean-fields being projected upon. In accordance with this approach we use the Undecimated Discrete Wavelet Transform (UDWT) method to decompose the fields into \emph{local scale-dependent} background magnetic fields and fluctuations, $\overline{\mathbf{B}}(t,f)$ and $\delta\mathbf{B}(t,f)$ respectively; where $f$ explicitly shows the frequency or scale dependence -- the parallel and transverse fluctuations, $\delta \mathbf{B}_{(\parallel/\perp)}(t,f)$, are then obtained from these.The background to this particular wavelet method and brief details of the algorithms are described in Appendix \ref{sec:UDWT}.


\section{Results and Discussions} \label{sec:Results} 
\subsection{Power isotropy and enhanced magnetic compressibility} \label{sec:CompIsotropy} 

\begin{figure}
	\begin{centering}
		\includegraphics[width=1\columnwidth]{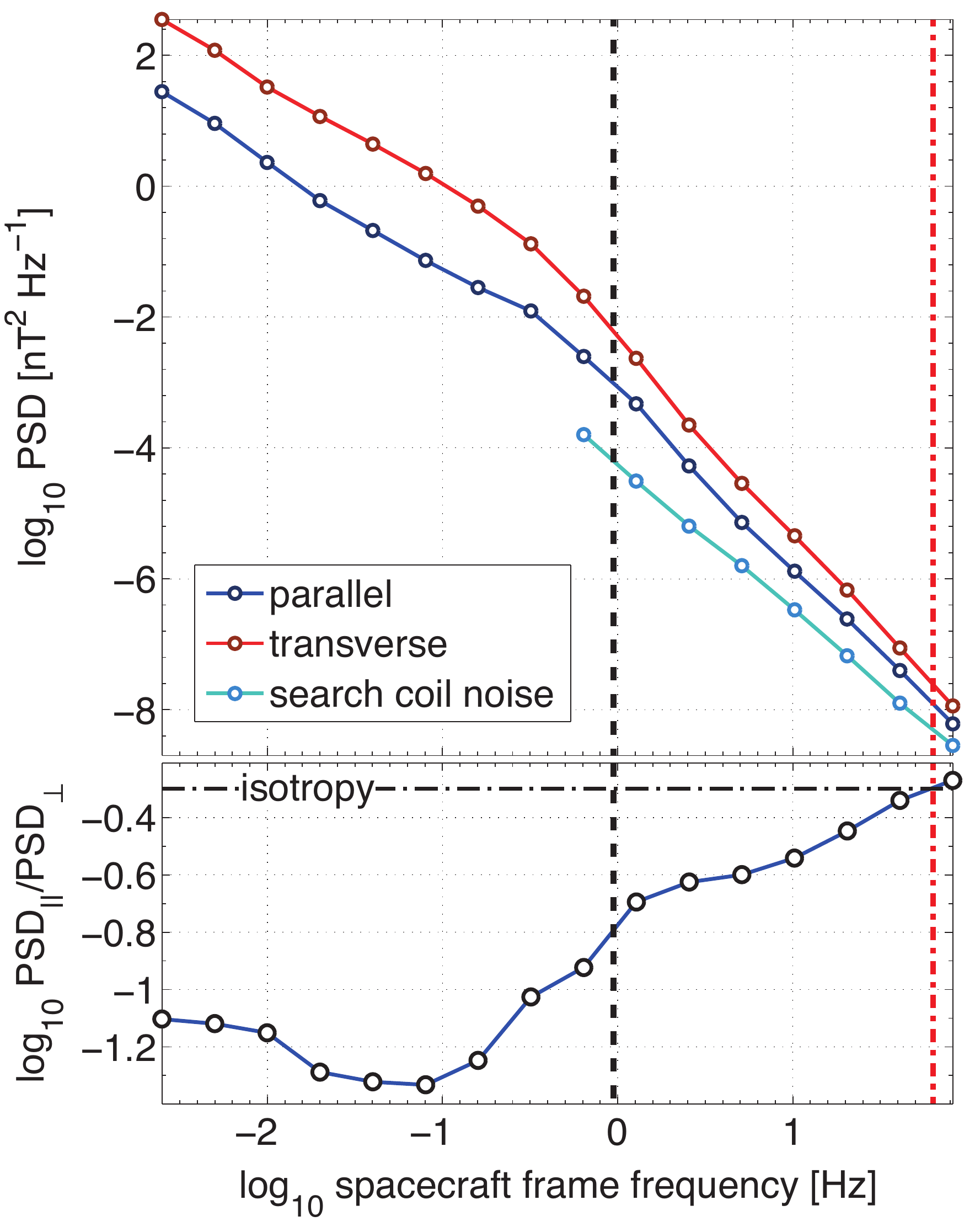} \par
	\end{centering}
	\caption{\label{fig:PSD}(Upper panel) PSD (from Cluster) of the transverse and parallel components spanning the inertial and dissipation ranges. Standardised sample size variance errors are smaller than the markers. The search-coil sensitivity floor PSD is obtained from the $z$-component (spacecraft SR2 coordinates) of a very quiet period in the magnetotail lobes (2007/06/30 15:00-15:05 UT) as a proxy for the instrumentation noise. (Lower panel) Ratio of parallel over transverse PSD. Horizontal dot-dashed line indicates a ratio of $1/3$ where isotropy in power occurs. Vertical dashed and dashed-dotted lines indicate $\rho_{i}$ and $\rho_{e}$ respectively, Doppler-shifted to spacecraft frequency using the Taylor hypothesis. The little blip in $PSD_{\parallel}$ at $\sim0.25$ Hz (and in $PSD_{\parallel}/PSD_{\perp}$) is due to the residual spacecraft ($\sim 4$ seconds) spin-tone in the FGM signal; it is more noticeable in $PSD_{\parallel}$ due to the lower power in parallel fluctuations in the inertial range.} 
\end{figure}

In keeping with Parseval's theorem for the conservation of the $L^{2}$-norm (energy conservation) the wavelet PSD for the parallel and transverse magnetic field components is given by
\begin{equation}
	PSD_{\parallel(\perp)}(f)=\frac{2\Delta}{N}\sum_{j=1}^{N}\delta B^{2}_{\parallel(\perp)}(t_{j},f)\ ,\label{eq:1} 
\end{equation}
where $\delta B_{\perp}(t_{j},f)=\sqrt{\delta B_{\perp1}^{2}(t_{j},f)+\delta B_{\perp2}^{2}(t_{j},f)}$ is the magnitude of the total transverse fluctuations at time $t_{j}$ and frequency $f$, $\Delta$ is the sampling period between each measurement, and $N$ is the sample size at each frequency $f$. For the Cluster interval the $PSD_{\parallel}$ and $PSD_{\perp}$ are shown in fig.\ref{fig:PSD}. The spectral indices obtained are $\simeq-1.62\pm0.01$ and $\simeq-1.59\pm0.01$ for parallel and transverse components respectively, in the inertial range; and $\simeq-2.67\pm0.01$ and $\simeq-2.94\pm0.01$ for parallel and transverse components respectively, in the dissipation range. The lower panel of fig.\ref{fig:PSD} shows that not only do these results recover the $\sim9:1$ anisotropy ratio of \citet{BelcherDavis1971}, they also show that the decrease in the anisotropy observed by \citep{Leamon1998b,Alexandrova2008c,Hamilton2008} in the dissipation range is actually a scale-free progression to isotropy. This progression of the anisotropy in the power ratio $PSD_{\parallel/\perp}$ (a measure of magnetic compressible fluctuations) begins at the spectral break (spacecraft frequency $\sim0.25$ Hz), just before the calculated $\rho_{i}$, and follows the power-law relationship $PSD_{\parallel/\perp}\sim f^{1/3\pm0.05}$ to $\rho_{e}$, where isotropy in power between the three components (parallel and two transverse components) is achieved. This isotropy corresponds to a value of $PSD_{\parallel}/PSD_{\perp}=1/2$ and is indicated in the lower panel of fig. \ref{fig:PSD}. Although this enhancement of parallel, or compressible, fluctuations in the dissipation range has already been commented upon by various authors \citep{Leamon1998b,Hamilton2008,Alexandrova2008c}, it is normally shown to be nearly constant (apart from in \citep{Salem2012}). To our knowledge this is the first time that an observation of \emph{isotropy} has been noted to occur at $k\rho_{e}\simeq1$; although it is also strongly suggested in the PSDs in \citet{Sahraoui2010}.

\begin{figure}
	\begin{centering}
		\includegraphics[width=1\columnwidth]{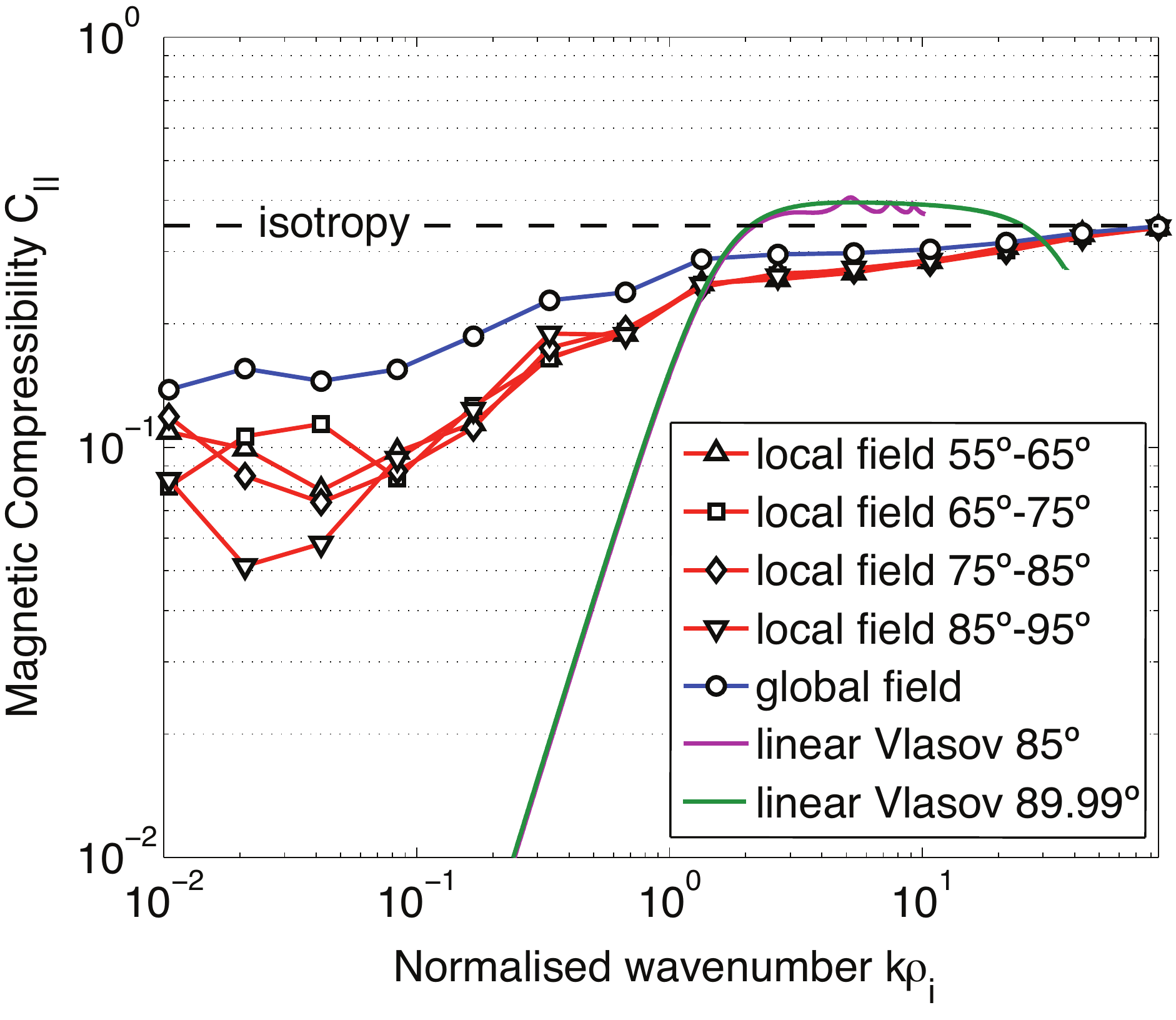} \par
	\end{centering}
	\caption{\label{fig:magcomp} Magnetic Compressibility $C_{||}$ (from Cluster), computed as in eqn. (\ref{eq:1.5}), against the normalised wavenumber for both local (scale-dependent) and global (computed over the whole interval) background magnetic fields. The normalised wavenumber was computed assuming the Taylor Hypothesis. Numerical solutions of the linearised Vlasov equation, corresponding to the kinetic Alfv\'en wave mode, (computed using the solar wind parameters for this interval) are also shown in comparison to illustrate the rising magnetic compressibility possessed by these modes at the ion-gyroscale. The horizontal dashed line at $C_{||}=1/3$ indicates the level at which full isotropy of the magnetic field fluctuations occur. As in fig. (\ref{fig:PSD}) the statistical sample size variance errors are smaller than the markers, and the effect of the spacecraft spin is clearly seen at $k\rho_{i}\sim0.25$.} 
\end{figure}

The lower panel of fig. \ref{fig:PSD} is calculated from the ratio of the averages of the parallel and transverse fluctuations to show a measure of the anisotropy. As such it does not constitute a proper estimate of the ensemble average of the anisotropy. A proper measure would be to take the ratio of $\delta B^{2}_{\parallel}(t_{j},f)$ and $\delta B^{2}_{\perp}(t_{j},f)$ at each time $t_{j}$, and then average over this ensemble of realisations of the anisotropy. However, this is prone to large errors induced by very large spikes caused by purely, or nearly pure, parallel fluctuations which result in divisions by a very small number. To overcome this problem and to obtain a proper ensemble averaged anisotropy measure, we compute an alternative and robust metric of the magnetic compressibility (similar in form to the expressions in \citep{Gary2009a, Alexandrova2008c}) defined as
\begin{equation}
	C_{\parallel}(f)=\sum_{j=1}^{N}\frac{\delta B^{2}_{\parallel}(t_{j},f)}{\delta B^{2}_{\parallel}(t_{j},f) +\delta B^{2}_{\perp}(t_{j},f)}\ ,\label{eq:1.5} 
\end{equation}
i.e. the compressibility is calculated locally from the time-dependent fluctuations and then averaged. Converting the spacecraft frequency into wavenumber using the Taylor hypothesis and normalising by the averaged  ion gyro radius for the interval, figure \ref{fig:magcomp} shows $C_{\parallel}(k\rho_{i})$ computed using the local scale-dependent mean field. 
We have binned the latter into $10^\circ$ angle bins (angle between $\bold{V}_{sw}$ and $\bold{e}_{\parallel}(t_{j},f)$) to show which components of the $\mathbf{k}$ variation we are measuring with respect to the local background (scale-dependent) magnetic field. On the same plot, and for comparison, we have also included the calculation of the magnetic compressibility using a global background magnetic field -- the latter consists of the mean average of the magnetic field vector over the whole interval being studied. The angle of the solar wind velocity to the background magnetic field using this global mean field would correspond to wavevector components at an angle of $\sim75^\circ$ to the background field, both within the inertial and dissipation ranges. From these plots of $C_{\parallel}(k\rho_{i})$ we can see that there is not only a large difference between $C_{\parallel}$ calculated using a local as opposed to a global field, there is also no significant difference between $C_{\parallel}$ calculated using a local field for the separate $\mathbf{k}$-component angle bins. Also, similar to the lower panel of fig. \ref{fig:PSD}, in all the curves we again see the enhancement of the magnetic compressibility -- with isotropy between all three components of the vector magnetic field fluctuations being reached at $k\rho_{e}\simeq1$.

Lastly, one should note from fig. \ref{fig:magcomp} that the difference between the magnetic compressibility computed with global and local scale-dependent background magnetic fields decreases as we move to smaller scales. At a first glance this is a non-intuitive result, as we would expect that the local field increasingly mimics the global field at larger scales. However, this can be easily resolved by the fact that if the fluctuations are isotropically distributed, it does not matter which basis one is projecting the fluctuations in -- they will look the same in all bases. So if the fluctuations are becoming more isotropic towards the smaller scales, the less will be the difference between the global and local background magnetic field bases.

\subsubsection{Rising magnetic compressibility in kinetic Alfv\'en wave solutions of the linearised Vlasov equation} \label{sec:kAws} 

Theories of plasma turbulence which advocate that a turbulent energy cascade within the dissipation range is mediated by the various linear wave modes of a plasma \citep{Leamon1999,Schekochihin2009,Gary2008}, suggest predictions for $C_{\parallel}$ \citep{Gary2009a,Sahraoui2012}. As the wave modes advocated to cascade energy at these scales are dispersive in nature (kinetic Alfv\'en and Whistler waves), many of the proponents of such theories suggest that what has been  dubbed the dissipation range for purely historical reasons, should actually be called the \emph{dispersive range} \citep{Stawicki2001,Sahraoui2012}. Within the context of sub-ion Larmor scales, there is a growing body of work showing that the fluctuations at these scales share the characteristic of kinetic Alfv\'en waves \citep{Leamon1998b,Leamon1999,Bale2005,Howes2008b,Sahraoui2009,Sahraoui2010,Salem2012}. Indeed, the enhancement of the magnetic compressibility seen above is strongly consistent with the transition, around $k\rho_{i}\sim1$, of purely incompressible shear Alfv\'en waves into kinetic Alfv\'en waves, with a strong compressive component \citep{Hollweg1999} and propagating at highly oblique (near-perpendicular) angles to the background magnetic field. 

In figure \ref{fig:magcomp} we overlay the observational results of $C_{\parallel}$ by predictions of $C_{\parallel}(k\rho_{i})$ from numerical solutions of the linearized Vlasov-equation for the kinetic Alfv\'en wave mode at a highly oblique ($85^\circ$) angle and at a virtually perpendicular ($89.99^\circ$) angle, using the same bulk plasma field and particle parameters of our data. These numerical solutions were plotted using the WHAMP code \citep{Ronnmark1982} which assumes a strong guide field i.e. a global background magnetic field. The plot shows that our results are qualitatively consistent with $C_{\parallel}$ predicted for kinetic Alfv\'en waves. This enhancement of the compressive component in the kinetic Alfv\'en wave is due to the coupling of the Alfv\'en mode to the purely compressive ion acoustic mode \citep{Leamon1999}. In the inertial range below $k\rho_{i}\sim1$ both of these modes are decoupled \citep{Howes2006,Schekochihin2009}. In the context of such theories, the 10\% compressible component that we see in our results in the inertial range is most likely due to the presence of magnetosonic compressible modes in the data, all at highly oblique (nearly perpendicular) angles to the background magnetic field. The modes can be in the form of slow modes, fast modes or non-propagating ($\omega=0$) pressure balanced structures advected by the flow (mirror modes), all of which have purely parallel (compressible) fluctuations \citep{Sahraoui2012}. Although a recent paper by \citet{Howes2011a} shows that slow modes or pressure balanced structures constitute the dominant part of the compressible fluctuations in the inertial range, fast magnetosonic modes  may also exist. This can be supported by the fact that these fast modes are shown to be considerably undamped (compared to slow modes) in the linear theory at highly oblique angles of propagation and high plasma beta ($\beta_{i}\gtrsim1$) \citep{Sahraoui2012}, conditions very relevant to the Solar Wind. In figure \ref{fig:magcomp}, the departure of our results from the theoretical solution at $k\rho_{i}<1$ can thus possibly be explained by the fact that the theoretical curve is only for kinetic Alfv\'en waves -- an addition of the above purely compressible magnetosonic modes might be able to reduce the difference significantly. In addition the theoretical plot is calculated assuming a strong global background magnetic field. Importantly, the theoretical plot for kinetic Alfv\'en waves shows that the magnetic compressibility should start to rise before $k\rho_{i}\sim1$ which is also what our results show. Recently, \citet{Salem2012} showed a similar result for magnetic compressibility (their definition of $C_{\parallel}$ not being a squared quantity as in this article) with better agreement with the theoretical $C_{\parallel}$ curve for kinetic Alfv\'en waves than the results presented here. The differences in our results and those of \citet{Salem2012} could be due to the intervals analysed: our interval is fast wind with $\beta_{i}\sim1.5$, whereas their interval is slow wind with $\beta_{i}\sim0.5$; that they used two-fluid linear warm plasma relations, instead of the linear Vlasov solutions used here; or that their component fluctuations were projected onto a global background magnetic field instead of a local scale-dependent one. To put both these results in their proper place, requires a large data survey with intervals that sample more broad and comprehensive plasma conditions.

\subsubsection{The role of the Hall term in the enhancement of magnetic compressibility and component isotropy} \label{sec:Hall} 

In the more general non-linear setting, this enhancement of magnetic compressible fluctuations can be explained by the increased prominence of the Hall term, $\mathbf{j}\times\mathbf{B}$, in the generalised Ohm's Law, where $\mathbf{j}$ is the current density arising from unequal ion and electron fluid velocities. To show this in the dynamical evolution of the magnetic field, we will turn to the Hall MHD model, as it is the simplest physical model of a plasma with unmagnetised ions \citep{Shaikh2009}. Although Hall MHD might not be an accurate representation of the full kinetic physics, especially the linear kinetic physics in hot plasmas \citep{Ito2004,Howes2009}, it serves as a simpler nonlinear model to explain the rising magnetic compressibility and the isotropy that we observe here. Primarily, we will look to the Hall-MHD induction equation (in SI units)
\begin{equation}
	\frac{\partial \bold{B}}{\partial t} =  \nabla\times( \bold{v} \times \bold{B} ) + \frac{1}{\mu e n_{e}}\nabla\times(\bold{B}\cdot\nabla\bold{B}) + \eta\nabla^{2}\bold{B} \ ,\label{eq:1.1} 
\end{equation}
where $\mathbf{v}$ is the bulk (ion) velocity, $\eta$ is the magnetic diffusivity, $\mu$ is the permeability, $e$ is the electron charge and $n_{e}$ is the electron number density. The first term on the RHS of equation (\ref{eq:1.1}) is the convective/dynamo term, the second term is the Hall term and the last term is the diffusive term. We have neglected the effects of any electron pressure gradients here to simplify the discussion, and as these are $\mathcal{O}(\delta n_{e}/n^{2}_{e})$ their effects are considered small. We have kept the diffusive term -- which is normally negligible for these solar wind parameters -- in order to retain an energy sink in the equations, but also for the possibility of a renormalized, or turbulent, magnetic diffusivity; however, we will not be making use of this term in the following arguments. Although other terms, such as the electron inertia that are very relevant to high frequency modes e.g. Whistler modes, become relevant at scales $k\lambda_{e}\sim 1$ (where $\lambda_{e}$ is the electron inertial length), we stick to the Hall MHD model above and focus on the effects of the Hall term with respect to the convective term.  Importantly, in $\beta \sim \mathcal{O}(1)$ plasmas ($\lambda_{i}\sim\rho_{i}$ as $\lambda_{i}=\rho_{i}/\sqrt{\beta}$), such as the solar wind at 1AU, as one approaches the ion-Larmor scale $\rho_{i}$, the Hall term becomes of the order of the convective term \citep{Goossens2003}. In the inertial range, at scales above the Larmor radius, it is the convective term which dominates and the magnetic field is frozen to the ion flow, such that any flow of the plasma \emph{perpendicular} to the magnetic field affects the evolution of the magnetic field. Thus, if we assume that much of the power in inertial range magnetic field fluctuations is locally generated from the velocity fluctuations (via the non-linear evolution of the ion momentum equation) in the form of a dynamo effect ($\nabla \times (\mathbf{v} \times \mathbf{B})$ term in (\ref{eq:1.1})), and is not being passively advected in the solar wind from the Sun, then it is natural that most of the power in the inertial range will be in transverse fluctuations -- also considered a signature of Alfv\'enic fluctuations \citep{BelcherDavis1971, Horbury2005}. Any residual parallel (compressible) fluctuations arise from fluctuations in the plasma density (which also contribute to the transverse fluctuations). 

If we now approach scales close to $\rho_{i}$ the Hall term starts to show its effects. This manifests itself in changing the direction of the fluctuations such that if transverse fluctuations dominate, the Hall term will generate parallel fluctuations and thus result in an increase in the magnetic compressibility $C_{\parallel}$. To see this we will focus on the Hall term and assume an oversimplification that a.) it is entirely transverse fluctuations $\delta B_{\perp}$ which dominate the turbulent fluctuations coming from the inertial range and b.) that $k_{\perp}$ fluctuations dominate (i.e. perpendicular gradients $\nabla_{\perp}$) -- both of which are supported by our observations, and studies of the full 3D $\bold{k}$-spectrum \citep{Sahraoui2010}. This results in the Hall term contribution to the induction equation becoming
\begin{equation}
	\left.\frac{\partial \bold{B}_{\parallel}}{\partial t}\right|_{Hall}\sim \nabla_{\perp 1}\times(\bold{B}\cdot\nabla{\delta \bold{B}_{\perp 2}}) + \nabla_{\perp 2}\times(\bold{B}\cdot\nabla{\delta \bold{B}_{\perp1}})\ ,\label{eq:1.2} 
\end{equation}
where we have explicitly separated the two transverse components to make clear the that the two directions are not parallel to each other in the vector cross product.  Equation (\ref{eq:1.2}) clearly shows the origin of the enhancement in the parallel magnetic compressible fluctuations. In $k$-space the RHS of equation (\ref{eq:1.2}) would be in the form of the following convolution
\begin{equation}
	\int d^{3}\bold{j}\ \bold{k}\cdot\bold{B}(\bold{k}-\bold{j},t)(\bold{k}_{\perp 1} \times \delta\bold{B}_{\perp 2}(\bold{j},t) + \bold{k}_{\perp 2} \times \delta\bold{B}_{\perp 1}(\bold{j},t)),\label{eq:1.3} 
\end{equation}
where the wavevector arguments are shown explicitly. More importantly, equations (\ref{eq:1.2}) and (\ref{eq:1.3}) also suggest the origin of the isotropy that we observe amongst the different fluctuation components at scales close to $\rho_{e}$, where terms similar to (\ref{eq:1.2}) and (\ref{eq:1.3}) are dominant and the convective term becomes negligible. At such scales, the Hall term will dominate and will also convert any parallel fluctuations which are generated (or already present) into transverse ones. This will evolve until a steady state is reached between the various terms corresponding to the vector components of the magnetic field fluctuations i.e. isotropy between all the vector components. Interestingly, this simple observation of the role of the Hall term in the enhancement of magnetic compressibility and the resultant isotropy, also suggests some information on the $k$-anisotropy  i.e. $k_{\parallel}/k_{\perp}$. If, say, equation (\ref{eq:1.3}) was dominated by $k_{\parallel}$ instead of $k_{\perp}$, then it is clear that an inertial range dominated by transverse fluctuations will transition into something which is also dominated by transverse fluctuations, with little or no parallel fluctuations -- this is not what we observe here. This indicates that, in our Hall-MHD formalism, power in $k_{\parallel}$ fluctuations is not dominant. Intriguingly, however, this type of argument also does not rule out the possibility that \emph{both} $k_{\parallel}$ and $k_{\perp}$ fluctuations are significant at sub-Larmor scales \citep{Gary2010}; the only observational study of the full 3D $P(\bold{k})$ spectra at such scales by \citet{Sahraoui2010}, seems to indicate that this is not the case. This and more detailed calculations of the energy transfer between wavevector triads of fluctuations will be the subject of a future investigation, so we can determine a more precise nature of such steady states. Crucially, measurements of the magnetic compressibility, in the context of a Hall-MHD model, could possibly constrain the measurements of the $\bold{k}$ vector anisotropy. 

\subsubsection{The transition range} \label{sec:Transition} 

A \emph{transition range} between the inertial and dissipation ranges was suggested by \citet{Sahraoui2010}, which was, in the context of dispersive waves, cited as a possible sign of where magnetic field energy is Landau damped into ion heating. This transition range, comprising just under a decade of scales around $k\rho_{i}\sim1$ is also seen in the reduced spectra in \citep{kiyani2009b,Chen2010b} and is distinctly different from the power-law which follows at smaller scales down to $k\rho_{e}\sim1$ (see fig.1 in \citep{kiyani2009b} and fig.2 in \citep{Sahraoui2010}). Our arguments above suggest that the competition between the convective and Hall terms could also explain such a transition range. In this case, and from looking at where the compressible fluctuations begin to rise in fig.\ref{fig:PSD} and fig.\ref{fig:magcomp} ($k\rho_{i}\sim0.1$), it seems that the transition range could actually span a much larger range of scales. It is important to note here that we do not know what the functional form of the transition from the convective dominated regime to the Hall dominated regime is. All we know is that in the inertial range, far from the spectral break, one can neglect the effects of the Hall term; well below the ion-Larmor radius and well past the spectral break we can neglect the convective term; and at $k\rho_{i}\simeq1$ (for $\beta_{i}\sim1$) these terms are of the order of each other \citep{Goossens2003}. Using the above arguments, and the latter observation about the similar strengths of the convective and Hall terms at $k\rho_{i}\simeq1$ we can make the following observation: at $k\rho_{i}\simeq1$ the convective term will provide half of the fluctuation power (50\%) which we assume, from the arguments and observations above, will be comprised of 10\% fluctuations in the parallel direction; the Hall term will contribute the other half of the fluctuation power (50\%), and 33\% of these fluctuations will be in the parallel direction. This means that the magnetic compressibility at $k\rho_{i}\simeq1$ will be 
\begin{equation}
C_{\parallel}\simeq 0.5\times0.33 +0.5\times 0.1\simeq0.22\ .\label{eq:1.4}
\end{equation}
This value is in excellent agreement with the value of $C_{\parallel}$ at $k\rho_{i}\simeq1$ extrapolated from the local field curves in fig.\ref{fig:magcomp}.

\subsection{Higher-order statistics and intermittency} \label{sec:HoS} 
\begin{figure}
	\begin{centering}
		\includegraphics[width=1\columnwidth]{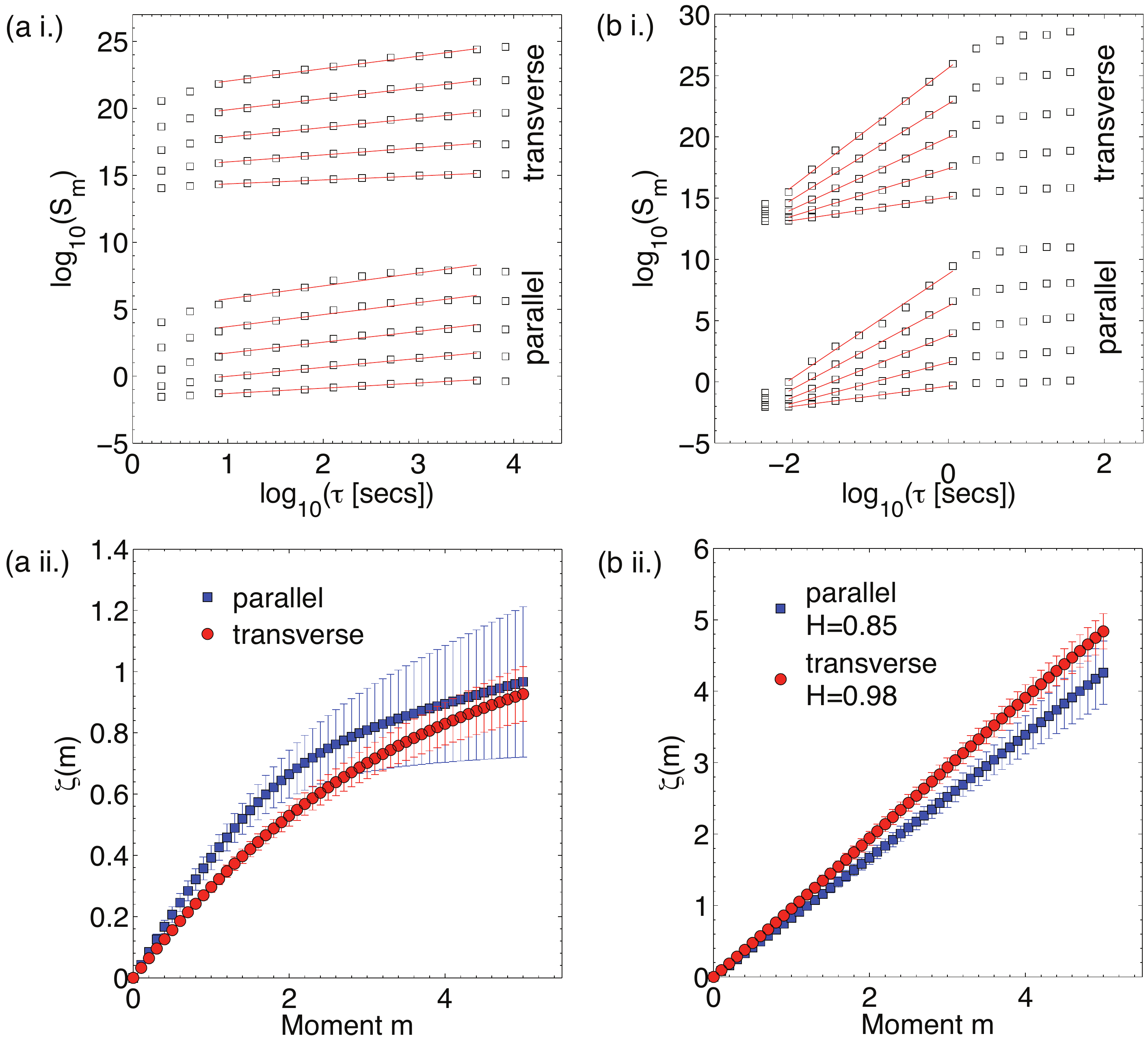} \par
	\end{centering}
	\caption{\label{fig:zeta}(a i.) Transverse and parallel wavelet structure functions of order $1-5$ (from the bottom) and (a ii.) resultant scaling exponents for the inertial range using the ACE data interval. Structure functions have been vertically shifted for clarity. (b i.) and (b ii.) have similar descriptions to (a i.) and (a ii.) but are from data in the dissipation range using the Cluster data interval. These are the anisotropic generlisations of the differences in the scaling behaviour between the inertial and dissipation ranges first shown by \citet{kiyani2009b}.} 
\end{figure}

We next calculate higher-order statistics given by the structure functions (absolute moments of the fluctuations) \citep{kiyani2009b} for the different components of the magnetic field fluctuations with respect to the local scale-dependent background magnetic field. The $m^{th}$ order wavelet structure function \citep{Farge2006} is given by
\begin{equation}
	S_{\parallel(\perp)}^{m}(\tau)=\frac{1}{N}\sum_{j=1}^{N}\left|\frac{\delta B_{\parallel(\perp)}(t_{j},\tau)}{\sqrt{\tau}}\right|^{m}\ , \label{eq:2} 
\end{equation}
where, as detailed in Appendix \ref{sec:UDWT}, $\tau=2^{i}\Delta:\ i=\left\{ 0,1,2,3,\ldots\right\} $ is the dyadic time scale parameter related to the central frequency $f$ used earlier, and $\Delta$ is the sampling period. Scale-invariance is indicated by $S_{\parallel(\perp)}^{m}(\tau)\propto\tau^{\zeta(m)}$; where $\zeta(m)$ are the scaling exponents. The structure functions and corresponding scaling exponents $\zeta(m)$ are shown in fig. \ref{fig:zeta} for both the inertial and dissipation ranges using the ACE and Cluster intervals respectively. 

Similar to the results by \citet{kiyani2009b}, the higher-order scaling in the inertial and dissipation ranges are distinct. The inertial range shows multi-exponent scaling as evidenced by a nonlinear $\zeta(m)$ characteristic of solar wind turbulence at MHD scales \citep{TuMarsch1995}. In contrast, the dissipation range is monoscaling i.e. characterised by a linear $\zeta(m)=Hm$ and a single exponent $H$. Notably both parallel and transverse fluctuations in the dissipation range show this different scaling behaviour in the inertial and dissipation ranges. However, the scaling behaviour between the different components is also distinct; with a more pronounced difference shown in the inertial range. The difference in the dissipation range exponents for parallel and transverse fluctuations is simply reflecting the different spectral exponents seen earlier in fig.\ref{fig:PSD}.

\begin{figure}
	\begin{centering}
		\includegraphics[width=1\columnwidth]{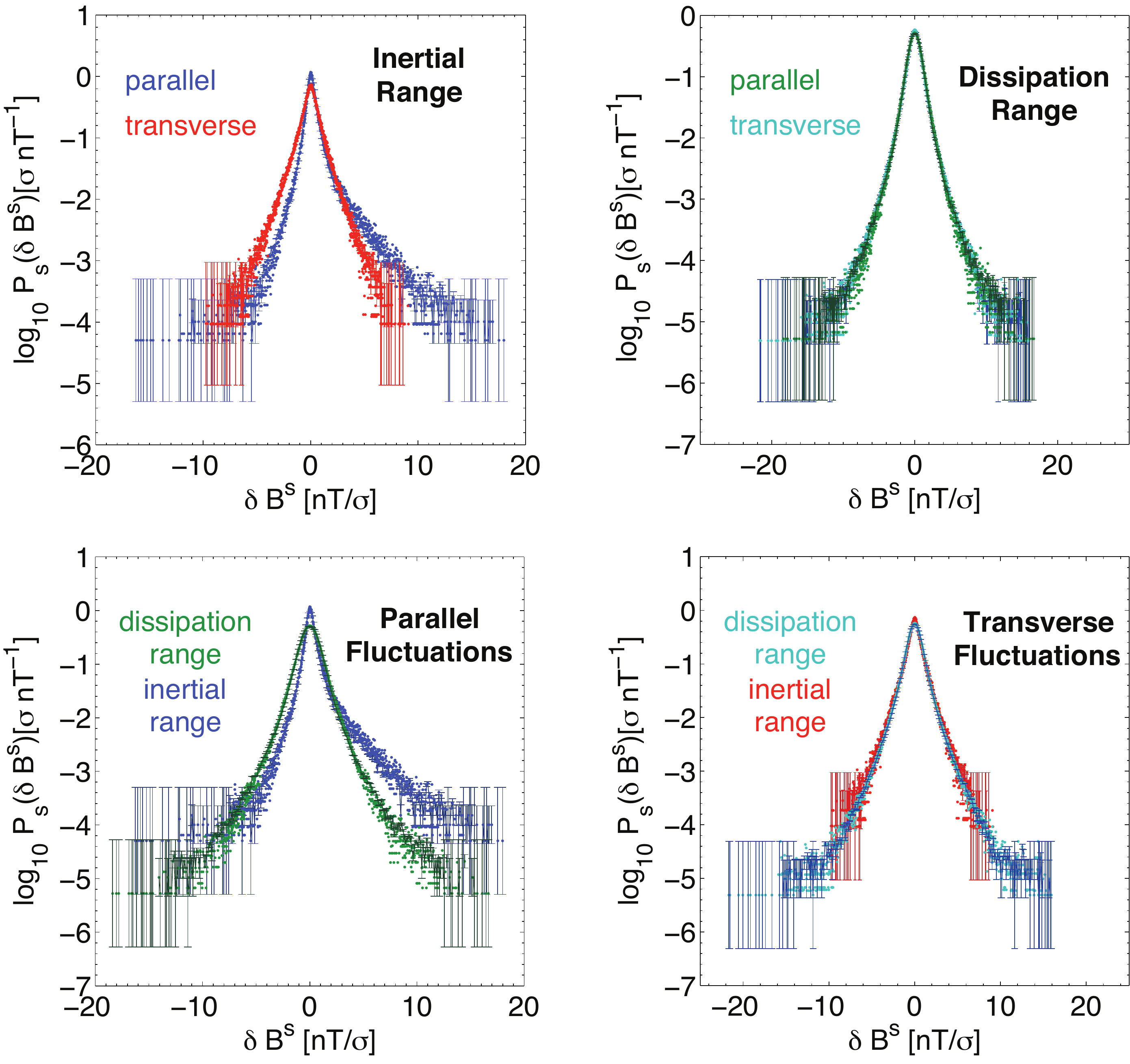} \par
	\end{centering}
	\caption{\label{fig:PDF}Rescaled (standardised) transverse and parallel PDFs of fluctuations in the inertial and dissipation ranges (with Poisonnian error bars). Normalised histograms with 300 bins each were used to compute the PDFs. Four values of $\tau$ were used in both ranges: $\tau=\left\{ 16,\ 32,\ 64,\ 128\right\}$ seconds for the inertial range from ACE, and $\tau=\left\{ 0.036,\ 0.071,\ 0.142,\ 0.284\right\}$ seconds for the dissipation range from Cluster.} 
\end{figure}

Before we discuss these higher-order scaling results, we complete the statistical results by finally looking scale-by-scale at the individual probability density functions (PDFs) for the transverse ($e_{\perp1}$ direction) and parallel fluctuations. It is necessary to pick one of the transverse directions as the combined (magnitude of) transverse fluctuations are positive-definite quantities, and thus do not illustrate any symmetric/asymmetric character of the fluctuations. There is no \emph{a priori} reason, from the symmetry of the physics, for one not to expect the fluctuations to be axisymmetrically distributed in the plane perpendicular to the background magnetic field. However, if we look at the separate PSDs for the two transverse components of the magnetic field fluctuations (not shown here) we will notice that this symmetry is broken and the power in the two transverse components is distinctly different. When we constructed our scale-dependent orthonormal bases (Appendix \ref{sec:UDWT}) it was natural to involve the background guide field as it is ubiquitously known to order the physics in magnetised plasmas. The other two directions perpendicular to this are relatively arbitrary, and we naturally chose the stable mean bulk velocity field direction, $\bold{\hat{V}_{sw}}$, to form these in the manner of \citet{BelcherDavis1971}: an orthonormal scale-dependent `field-velocity' coordinate system. However, the solar wind bulk velocity field picks a preferred sampling direction (in \emph{k}-space) resulting in the reduced spectrum mentioned earlier in the Introduction. In introducing the velocity field in such a way, the reduced spectrum breaks the transverse axisymmetry of the magnetic field fluctuations and introduces a measurement bias (see also \citet{Turner2011} who use comparisons with MHD turbulence simulations, and a model of transverse waves to show the importance of the spectral slope in this broken symmetry). In our scenario where we have taken a constant velocity field direction, $\bold{\hat{V}_{sw}}$, throughout our entire interval (due to the dominance of the GSE $x$-component), and the background magnetic field remains mostly highly oblique to $\bold{\hat{V}_{sw}}$, this bias in the reduced spectra remains nearly the same at all times, and effectively results in a simple (constant) shift in the (log) frequency which makes the two transverse PSDs misalign with each other -- it is essentially a Doppler shift-like sampling effect. The magnitude of the transverse vector is not affected by such a bias, so all our results above are ambivalent to this symmetry breaking. Although the PSD of the two transverse components differ in this respect, we can confirm that the standardised (rescaled) PDFs for both transverse components are near-identical. Thus, the breaking of the transverse axisymmetry does not affect the results presented in this paper.

The PDFs for $\delta B_{\parallel}$ and $\delta B_{\perp1}$ are shown in fig. \ref{fig:PDF}, where we have used the self-affine scaling operation $\mathcal{P}_{s}(\delta B_{i}\sigma^{-1})=\sigma P\left(\delta B_{i},\tau\right)\ $ to rescale (standardise) the fluctuations by their standard deviation so to offer a comparison of the functional form of the PDFs for different $\tau$. We show overlaid rescaled PDFs corresponding to four values of $\tau$ in the dissipation range, $\tau=\left\{ 0.036,\ 0.071,\ 0.142,\ 0.284\right\}$ seconds (from Cluster), and in the inertial range, $\tau=\left\{ 16,\ 32,\ 64,\ 128\right\}$ seconds (from ACE). Figure \ref{fig:PDF} shows that in the dissipation range \emph{the same} PDF functional form is obtained over the range of $\tau$ for both parallel and transverse fluctuations -- suggesting that the dissipation range is in this sense `process isotropic' with a (speculatively) single physical process determining the dynamics. This supports our arguments and results above which show the growing isotropy in the components of the magnetic field fluctuations. This is in contrast to the inertial range where we see that the transverse and parallel PDFs are clearly different as we would anticipate from their distinct $\zeta(m)$ shown in fig. \ref{fig:zeta} (a ii). 

\begin{figure}
	\begin{centering}
		\includegraphics[width=1\columnwidth]{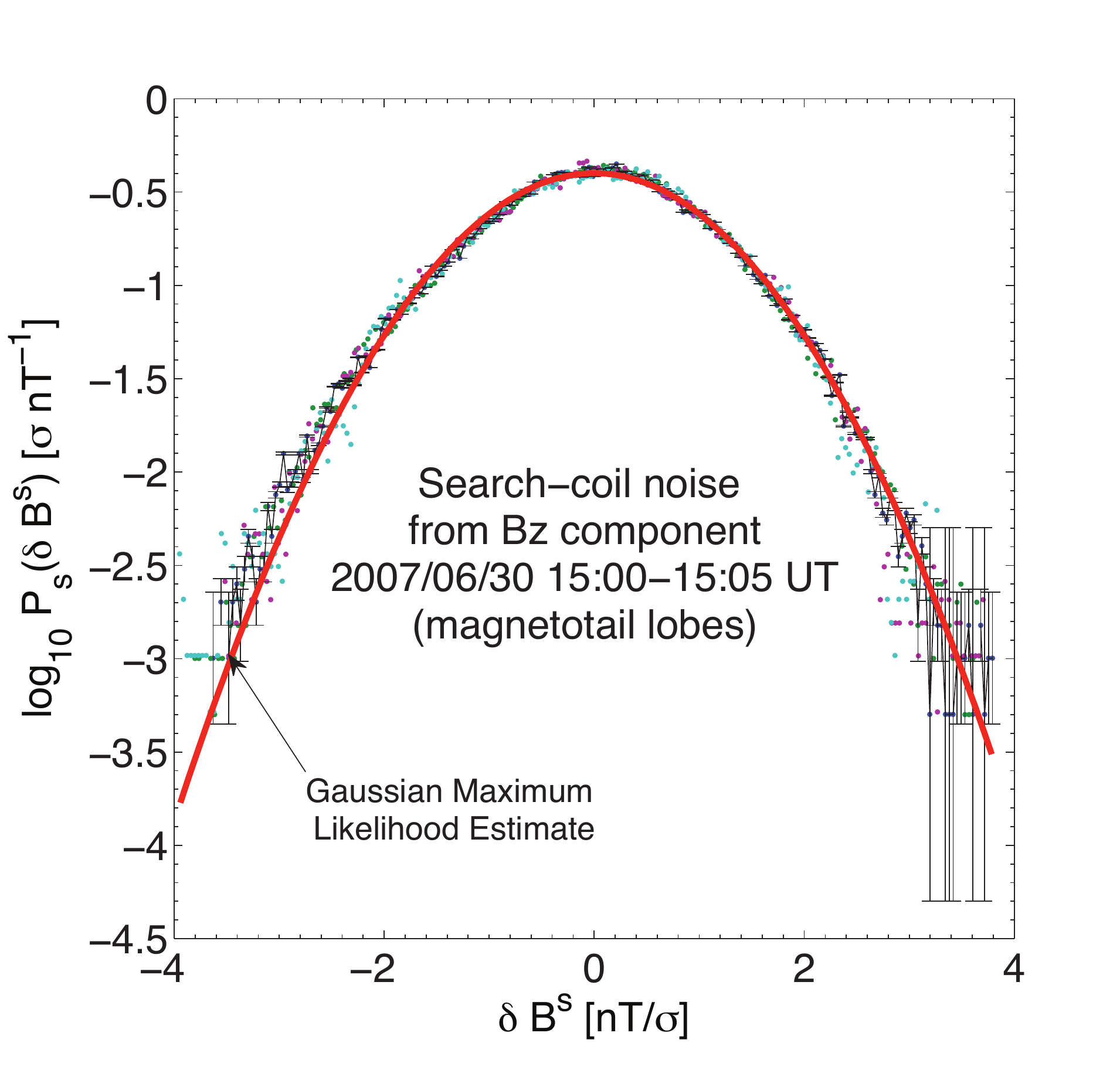} \par
	\end{centering}
	\caption{\label{fig:noise}Rescaled PDF of $B_{z}$ magnetic fluctuations from the instrument noise proxy of fig. \ref{fig:PSD}, at the same scales $\tau$ as in fig. \ref{fig:PDF} for the dissipation range. A Maximum Likelihood Estimate clearly shows the Gaussian nature of this noise proxy -- very different from the highly non-Gaussian statistics shown in fig.\ref{fig:PDF}. We can also confirm that the smallest scales in our data which coincide with the noise floor, in terms of power and Gaussian statistics, are excluded from the study presented in this article.} 
\end{figure}

In addition to the differences in the scaling behaviour, both inertial and dissipation range PDFs are highly non-Gaussian indicating high levels of intermittency. In contrast, the rescaled PDFs of the STAFF-SCM instrument noise shown in fig. \ref{fig:noise} is clearly Gaussian; confirming that our results are robust to noise contamination. Furthermore, we have analysed other intervals, 2007/20/01 12:00-13:15UT and 2007/20/01 13:30-14:10UT, and find broadly consistent results with those presented above ($H_{\parallel}=0.78$, $H_{\perp}=0.95$ and $H_{\parallel}=0.8$, $H_{\perp}=0.94$ respectively), but with lower signal-to-noise ratios. Please note, we consider `intermittency' in the most general sense, to simply reflect the presence of rare, or `bursty', large amplitude fluctuations in our signals -- something which is manifest in our non-Gaussian PDFs in fig. \ref{fig:PDF}. This is in contrast to the popular definition, advocated by \citet{Frisch1995}, which links intermittency to a \emph{dependency} of the kurtosis $\kappa(\tau)=S^{4}(\tau)/(S^{2}(\tau))^{2}$ (or another higher order moment ratio) with scale $\tau$ -- a property of multifractal (multi-exponent) scaling i.e as seen for the inertial range in our results. In this latter definition, the constancy of the kurtosis with scale indicates `self-similar' signals such as our monoscaling signature seen in the dissipation range. There is no need to plot the kurtosis for our data as this can be very easily seen if we use the scaling relationship of the $m^{th}$-order structure function $S^{m}(\tau)=\tau^{\zeta(m)}S^{m}(1)$ (see \citep{kiyani2006} for notation) in the definition for the kurtosis. This will result in $\kappa(\tau)=\tau^{\zeta(4)-2\zeta(2)}\kappa(1)$ which, if $\zeta(m)$ is a linear function of $m$ ($\zeta(m)=Hm$), is independent of $\tau$; as is the case for our results in the dissipation range. However, even though our results show a monoscaling signature in the dissipation range, there is still a small nominal dependence on the scale $\tau$ for $\kappa(\tau)$. This could be simply due to a statistical artifact associated with finite sample size and the fact that the kurtosis is sensitive to very large fluctuations. Our very largest events at the tails of our distribution are not statistically well-sampled --  an un-avoidable pit-fall of heavy-tailed distributions, as seen by the large errors at the tails of our PDFs in fig.\ref{fig:PDF}. 

\subsubsection{Comparison with other works on scaling at $k\rho_{i}>1$} \label{sec:Comparison} 
 
We now compare our results for higher-order statistics in the dissipation range to other studies from both observations and simulations. In the case of the dissipation range, there are only really two of these: the study of \citet{Alexandrova2008c} who used normal-mode (25 Hz) Cluster observations, and \citet{Cho2009} who conducted $512^{3}$ Electron MHD (EMHD) simulations. Although EMHD is a high frequency model for magnetic field dynamics with inertialess ions (the opposite of Hall MHD), within the scales that we are analysing here, $\rho^{-1}_{i} \lesssim k \lesssim \rho^{-1}_{e}$, the induction equation for both Hall MHD and EMHD are near-identical for an order unity $\beta_{i}$ plasma. Thus the authors \citep{Cho2009} also use EMHD to study the dynamics and statistics of magnetic field fluctuations in solar wind turbulence.

\citet{Alexandrova2008c} used a continuous wavelet transform using the Morlet wavelet to calculate both PDFs and flatness (kurtosis) functions. They too found non-Gaussian PDFs within the dissipation range but, in contrast to our monoscaling results, showed that the dissipation range has a kurtosis which increases rapidly with scale $\tau$ i.e. suggestive of multifractal scaling. They also showed a dependence of this flatness function with $\beta_{i}$. The increase of compressible fluctuations coincided with a steepening of the spectral slope \emph{and} the increase of intermittency, as reflected in the flatness function. This prompted the authors to suggest that, another non-linear energy cascade, akin to the inertial range, takes place in the dissipation range. However, due to the increased role of compressible fluctuations at these scales, driven by plasma density fluctuations, this is a small-scale \emph{compressible cascade}. Due to the dominance of the Hall term at these scales, \citet{Alexandrova2008c} then suggest a simple model based on compressible Hall MHD, which takes into account density fluctuations to describe the deviation of the energy spectrum slope from the $-7/3$ predicted purely by the induction equation for incompressible EMHD \citep{Biskamp1996}. Without a further study of more data intervals, it is difficult to say why our interval shows quite a different scaling behaviour to the interval investigated by \citet{Alexandrova2008c}. However, considering our results and our earlier arguments regarding the magnetic compressibility and the role of the Hall term, the notion of a compressible cascade is an appealing one. Although specific results were not explicitly shown, we should also mention the work of   \citet{Sahraoui2010b}, where the monoscaling result presented in this article and in \citep{kiyani2009b} was also independently confirmed with Cluster magnetic field data.

The very informative article by \citet{Cho2009} describes the results of their EMHD turbulence simulations, as well as ERMHD simulations. The UDWT digital filters used in our study are similar (in spirit) to the multi-point structure functions used in the analysis of \cite{Cho2009}. The main feature of these simulations which is relevant to our results is that, for the EMHD simulations, the monoscaling behaviour found in our results is also seen in the higher-order statistics computed from multi-point structure functions for moment orders up to $m=5$. Computing higher order moments greater than 5 becomes very difficult in non-Gaussian heavy-tailed statistics due to the poor sampling in the tails \citep{ddw2004}. The scaling exponent $H$ for their $\zeta(m)=Hm$ scaling (as they used $\zeta(p)/\zeta(3)$) is not explicitly stated in their paper. However, this is easily obtained from the  $-7/3$ spectral slope that they obtain in their energy spectrum (for both EMHD and ERMHD), using the well known relationship $\beta=-(2H+1)$ \citep{kiyani2009b}, where $\beta$ is the spectral slope. We can then infer that $H=2/3\simeq0.66$. This is the scaling for $\bold{r}_{\perp}$ ($\bold{k}_{\perp}$) variations. The results of \citet{Cho2009} also show moderately non-Gaussian PDFs. The authors then compute two-point flatness functions, as well as wavelet flatness functions using the Morlet wavelet similar to \citet{Alexandrova2008c}. Interestingly, for $\bold{r}_{\perp}$ they find a nominal dependence of the flatness with separation scale $r$ (equivalent to $\tau$) for the two-point flatness function; but a larger increase with $r$ for the wavelet flatness function. This last part offers slightly conflicting results when considering their multipoint structure function monoscaling for moments of order less than 5. Nevertheless, the results of \citet{Cho2009} offer the closest agreement to our scaling results presented in this article, and serve as a good comparison to an appropriate model for dissipation range fluctuations (EMHD induction equation as opposed to ideal MHD). Similarly to \citet{Alexandrova2008c}, we could speculate that the difference in our exponents, $H_{\perp}=0.98$ and $H_{\parallel}=0.85$ (and also the respective spectral exponents), and the $-7/3$ from EMHD could be due to the differing role of density fluctuations. However, in recent gyrokinetic simulations of anisotropic plasma turbulence, \citet{Howes2011b} show that steeper than $-7/3$ spectra are only obtained if the full kinetic physics is considered in terms of retaining a physical damping mechanism e.g. collisionless Landau damping, transit time damping etc. This is absent in the EMHD, ERMHD and Hall MHD descriptions (see also \citep{Howes2009}). The gyrokinetic description retains this missing physics and thus \citet{Howes2011b} claim that their simulations successfully recover the steeper scaling seen here and in \citep{kiyani2009b,Alexandrova2009,Sahraoui2010} and \citep{Chen2010b} for similar plasma parameters.

\subsubsection{Data uncertainties and errors} \label{sec:Errors} 

There is a small but important data caveat that one should mention with regards to these results. This concerns the noise floor of the STAFF search-coil magnetometer and the signal-to-noise ratio at small scales near the electron gyroradius. Due to the dyadic frequency spacing of the UDWT that we use, frequency resolution is poor. We pay this expense for the faster algorithms and the smoother spectra that we gain; the latter being preferred for broadband spectra where robust estimates of the spectral slope are needed. This gives the impression that the signal to noise ratio at the smaller scales is higher than it might actually be. In reality there exist some high power spikes in the noise-floor signal $>60$Hz which could contaminate the observations at the electron gyro-scale (see figure 1 in \citep{kiyani2009b}). These spikes are due to interference from electrical signals of the Digital Wave Processor (DWP) instrument onboard Cluster \citep{DWPSpikes2012}. The DWP signals are used to synchronise instrument sampling within the experiments of the Cluster Wave Experiment Consortium (WEC), which STAFF belongs to. These spikes only affect the last data point in figures \ref{fig:PSD} and \ref{fig:magcomp} and thus should not change our results and conclusions significantly. Intervals with higher power such as those in the slow solar wind stream would be needed for better estimates at these higher frequencies. In addition, the noise-floor that we use is not the actual noise-floor but a proxy for it, obtained from a very quiet period in the magnetotail lobes where the magnetic field power is very low. The actual noise-floor could be smaller. A discussion on the noise-response of the STAFF search-coil instrument and its sensitivity in different plasma conditions can be found in \citet{Sahraoui2011}.

We would also like to note that the PDF for the parallel fluctuations seen in fig. \ref{fig:PDF} for the inertial range from ACE, show an unusual skew towards positive fluctuations. This is an artifact for this particular interval of ACE magnetic field data due to large spike-drops (discontinuities) in the magnetic field magnitude $|\bold{B}|$. As these spikes arise in $|\bold{B}|$ and not in the components, this reflects itself more in the parallel component of the fluctuations rather than the transverse components as, in many cases, one can use $|\bold{B}|$ as a proxy for parallel fluctuations e.g. in \citet{Alexandrova2008c}. This is also reflected in the large errors for the parallel scaling exponents in the inertial range shown in fig. \ref{fig:zeta} (a ii.). However, as these spikes are primarily in the large rare events in the positive tail of the PDF, we believe that the differences in the PDFs of the transverse and parallel fluctuations within the inertial range still occur and our discussion above remains valid. 

\subsubsection{Discussion} \label{sec:Discussion} 

The isotropy that we observe between the power in the different fluctuation components at $k\rho_{e}\sim1$ is by no means a result which would apply to all plasma environments. Indeed, there is some strong observational evidence that the ion/proton plasma $\beta_{i}$ also has a strong role to play here. This was shown by \citet{Hamilton2008}, who repeated the inertial range analysis of \citet{Smith2006c} within the dissipation range and showed that the variance anisotropy followed the empirical power-law relationship $\delta B_{\perp}^{2}/\delta B_{\parallel}^{2}\equiv PSD_{\perp}/PSD_{\parallel}\sim\beta_{p}^{-0.56}$ in open field line intervals of the solar wind from the ACE spacecraft. However, these results were limited to 3Hz cadence and thus could not sample as much of the dissipation range as in this article. Also \citet{Hamilton2008} used a global field (mean field over the entire interval) rather than the local scale-dependent field used in this article. This same dependence of the magnetic compressibility on $\beta_{i}$ was also indicated in higher cadence (25 Hz) Cluster data in the work by \citet{Alexandrova2008c}, again suggesting that higher $\beta_{i}$ implies higher levels of $C_{\parallel}$ (these authors used $\delta |\bold{B}|$ as a proxy for $\delta B_{\parallel}$). The effect of varying $\beta_{i}$ is not mentioned explicitly in our arguments using the Hall term explained above, as we have neglected density fluctuations to simplify the discussion. However, purely from the induction equation point of view, $\beta_{i}$ is taken into account in dissipation range scales if we retain the effects arising from density fluctuations at scales $k\rho_{i}>1$ as in the formulation of Electron Reduced MHD (ERMHD) \citep{Schekochihin2009,Cho2009}. In ERMHD the explicit $\beta_{i}$ dependence is shown, assuming pressure balance, in the pre-factor of the $\nabla_{\perp} \times (\bold{B}\cdot\nabla \bold{B}_{\perp})$ term in the induction equation for $B_{\parallel}$. Thus, this addition of density fluctuations can generalise our arguments above which were essentially based on the Hall term alone. At higher or lower than unity $\beta_{i}$, this will break the isotropy of the fluctuation components seen in this article at $k\rho_{e}\sim1$, as an additional anisotropic source of magnetic compressibility will be introduced via the density fluctuations (see also \citep{Malaspina2010} for high cadence measurements of density fluctuations in the solar wind between $\rho_{i}^{-1}<k<\rho_{e}^{-1}$). The explicit $\beta_{i}$ dependance for the rising compressibility is also seen in the fluid version of the kinetic Alfv\'en wave solution of compressible Hall MHD where the compressible component to the linear solution goes as $\delta \bold{B}_{\parallel}\sim k \lambda_{i}(\beta_{i}+1+\sin^{2} \theta_{kB})/\sin \theta_{kB}$ \citep{Sahraoui2003b}, where $\lambda_{i}$ is the ion inertial length, and $\theta_{kB}$ is the angle between the wavevector $\bold{k}$ and the magnetic field vector $\bold{B}$. This solution assumes that $\sin \theta_{kB}\neq0$; thus it is not valid for purely parallel propagating Alfv\'en waves. However, the rising compressibility is clearly seen here as $k\rightarrow \lambda_{i}$.

Note that in most of the previous arguments we have been slightly cavalier in our discussion of field components. Our introduction of the local background magnetic field should not be looked upon lightly. It is clear that in cases where $\delta \bold{B} \ll \bold{B_{0}}$, where $\bold{B_{0}}$ is the global background magnetic field, both global and local field descriptions should have negligible differences. However, as we observed in the section describing higher-order statistics, it is not obvious that this is the case here, as the distribution of $\delta\bold{B}$ is highly non-Gaussian with very heavy tails. In our explanation of the rising compressibility in sub-Larmor scales using the Hall-term and also the lower levels of the compressibility in the inertial range using the convective term in the induction equation, we did not mention the spatial (and possibly temporal) gradients which arise due to the now locally varying background magnetic field. In a global DC background field, or very slowly evolving field, these gradients are null or negligible. If there is a large spectral separation between the scales of the background magnetic field and the fluctuations that we are projecting on it, then this conclusion could still be quite valid. However, as our UDWT method assures that no such large gap exists, the statistics we calculate will include the effects of such gradients. Further investigation and mathematical grounding of a local-field is of crucial importance, as the notion of using a local versus a global background magnetic field is intimately tied with the issue of whether we have a strong background guide field or a weaker more stochastic one (see the discussion by \citet{Schekochihin2009} and references therein for further details and exploration of this topic; and also the recent discussion by \citet{Matthaeus2012} on the stochastic nature of the local background field and its relationship with global background field statistics). This also brings into question the use of linear wave modes, such as the kinetic Alfv\'en wave, which are normally described as small amplitude perturbations on a static or slowly evolving equilibrium field. Although finite amplitude propagating nonlinear Alfv\'en waves are an exact (Els\"asser) solution to the ideal MHD equations; and similar arguments in the context of the Electron Reduced MHD equations are made by \citet{Schekochihin2009} for finite amplitude kinetic Alfv\'en waves; this topic also requires further exploration (see also the excellent discussion of this by \citet{Dmitruk2009}, who explore the importance of the strength of the mean background field using direct numerical simulations of the incompressible MHD equations).

One might question a conclusion of a single physical process or type of mediating dynamics (in the dissipation range) by the observation that both transverse and parallel components have slightly different scaling exponents and spectral indices. This can be reconciled by realising that there is no contradiction in the fact that the two components can have slightly different scaling exponents but have identical PDF functional forms. Take for example the case of fractional Brownian motion (fBm) which is a non-Markovian generalisation of Brownian Motion. Two fBm stochastic processes with different scaling exponents will still have the same Gaussian functional form for their fluctuations. In fBm the scaling exponents are simply an additional parameter representing the degree of statistical dependence between the fluctuations \citep{mandelbrot1983,samorodnitsky1994}.

If we look from the point of view of linear wave mode characteristics, the distinctness between inertial and dissipation range scalings could also be seen to reflect the decoupled and coupled nature of the wave modes in the inertial range and dissipation ranges respectively \citep{Leamon1999}. The decoupling of these modes in the inertial range also results in the decoupling between transverse and compressible magnetic field fluctuations \citep{Schekochihin2009}. In this picture the scaling in the dissipation range could then be described by non-linearly interacting kinetic Alfv\'en waves. In all these cases, as they involve linear wave modes, the non-linear cascade would then take the form of `critical-balance' type cascades as in the works of \citet{Schekochihin2009} and \citet{Howes2008b,Howes2011b}; or if one can show that the non-linearity is small, a `weak-turbulence' type description \citep{Galtier2002}. Although there is no justification for the latter in the inertial range, there might be a case for it in the dissipation range if the coupling between modes becomes weak and results in small amplitude fluctuations \citep{Rudakov2011}. In hydrodynamics described by the Navier-Stokes equations, this latter assumption can be valid if one uses a local Reynolds number (see \citep{Batchelor1953}), which is small in the high-\emph{k} regime (when the spectral amplitudes become small) well into the hydrodynamic dissipation range. This can then provide a suitably small parameter to base a convergent (or asymptotic) perturbation expansion. Of course, without concrete measurements of non-linearity, this is all speculation. However, it could possibly be an avenue for a future (maybe fruitful) investigation. 

An open question is why such a transition from multifractal scaling to self-similar monscaling occurs between the inertial and dissipation ranges. \citet{Kiyani2010} briefly discuss this using hydrodynamic analogies \citep{FrischVergassola1991,Chevillard2005} of a near-dissipation range where the increased effect of viscosity successively switches off the available scaling exponents of the multifractal field -- equivalent to narrowing the `$f(\alpha)$' spectrum in multifractal models . However, no such apparent analogies with a viscous `fractal-dampening' can happen in a near-collisionless solar wind. Also, unlike the hydrodynamic analogy this transition happens very fast as we cross $k\rho_{i}\sim1$; although a scaling analysis within the `transition range' might yet reveal an intermediate scaling behaviour which shows this successive `switching-off' of scaling exponents. Indeed, the steepening of the spectral slope in the transition region is strongly suggestive that such an intermediate scaling range could possibly exist. This is an exciting prospect and will be the subject of a future investigation. This latter investigation is not feasible with our UDWT technique as due to the dyadic spacing, it has poor frequency resolution -- a more detailed `fine-toothed comb' technique such as the use of wavelet packets \citep{Walden1998} or the continuous wavelet transform \citep{Horbury2008,Podesta2009} would be needed here.


\section{Conclusions} \label{sec:Conclusions} In this article we have performed a systematic scale-by-scale anisotropic decomposition of magnetic field fluctuations in the solar wind across both the inertial and dissipation ranges, from MHD to electron scales. We use a novel scheme based on discrete wavelet filters which self-consistently decomposes the magnetic fields into scale-dependent background and fluctuating fields. From the background fields we then construct a local orthonormal scale-depedent field-velocity coordinate system, on which we then project the fluctuations and calculate our statistical quantities. In summary, our main findings are:-
\begin{enumerate}
\item There exists a successive \emph{scale-invariant} reduction in the power ratio between parallel and transverse components as we move to smaller scales below the ion-Larmor radius $\rho_{i}$. Importantly, at $k\rho_{e}\simeq1$ with $\bold{k}$ components highly oblique/perpendicular to the background magnetic field, the fluctuations become isotropically distributed between the power in all three components (two transverse and one parallel) i.e. equipartition of magnetic energy between all components. 
\item By comparing with linear solutions of the Vlasov equation and computing the magnetic compressibility we show that the scale-invariant reduction in the power ratio is also qualitatively consistent with the transition to kinetic Alfv\'en wave solutions which rapidly develop magnetic compressible components at scales around $\rho_{i}$.
\item Using a Hall MHD model we provide a simple (non-linear) description of this new result in terms of how the Hall term is responsible for the rise of parallel (compressible) magnetic fluctuations and the resultant isotropy at scales around $\rho_{e}$. 
\item Higher order statistics and PDFs for parallel/transverse fluctuations reveal that the dissipation range is characterised by a single monoscaling statistical signature with an \emph{isotropic} distribution of fluctuations; thus supporting the above observations of isotropy. This is in contrast to the inertial range where the corresponding signature is anisotropic and multiscaling (multifractal).
\end{enumerate}

In the compressible Hall-MHD framework, at such scales the velocity field is expected to decouple from the evolution of the magnetic field and some researchers \citep{Servidio2007,Alexandrova2008c} have suggested that a new non-linear \emph{compressible cascade} is responsible for the new power-law regime in the PSD. In this context, we should also mention that all of the arguments presented in this paper do not include any discussion of the possibility of phenomena arising from the effects of (turbulent) reconnection \citep{Lazarian2012} or of the possible role of a turbulent or renormalized magnetic diffusivity -- of which either, or both, are purported to be important at such scales. Neither have we discussed the role of density fluctuations, which have been neglected in our arguments, but are shown to play a possibly significant role through the electron compressibility \citep{Gary2009a}. Future work would involve looking at the possible importance of such fluctuations as well as investigating the energy transfer between modes in Hall-MHD/Electron MHD theories.

In the context of theories of turbulence which invoke plasma wave modes as the mediators of the energy cascade, two candidate wave modes are classically suggested in the dissipation range: quasi-perpendicular kinetic Alfv\'en waves and quasi-parallel Whistler waves. As first suggested by \citet{Gary2009a} and shown in \citet{Salem2012}, the magnetic compressibility, $C_{\parallel}$, might be a more robust measurement to distinguish between the different wave modes possibly mediating the energy cascade, than the phase speed plots calculated by $E/B$ ratios \citep{Bale2005,Sahraoui2009}. In hot plasmas ($\beta_{i}\gtrsim2$) and at highly oblique angles to the background magnetic field, classical Whistlers arising as an extension of the fast magnetosonic modes at high frequencies $\gtrsim\omega_{ci}$ (where $\omega_{ci}$ is the ion cyclotron frequency), are strongly damped and split up into different ion Bernstein modes. Recently, however, \citet{Sahraoui2012} have shown that, under the same condition, a Whistler-like branch having similar properties (e.g. dispersion and polarisation) as the classical Whistler mode, rises from the kinetic Alfv\'en wave mode and extends it to very high frequencies $\omega\gg\omega_{ci}$. Also, if there is a significant amount of power in shallow angles, e.g. $30^{\circ}$, between $\bold{k}$-vectors and the background magnetic field, then the classical quasi-parallel Whistlers could also show the same amount of magnetic compressibility that we observe in this paper \citep{Gary2009a,Salem2012}. 

Our study of higher order statistics show a new and significant result of the change of anisotropy at $\rho_{i}$. As in the behaviour of the inertial range, these scalings and statistics are a distinctive and important characteristic of the turbulence at sub-ion Larmor scales and any cascades which take place there. Therefore it is essential that any comprehensive statistical theory of plasma turbulence for such scales, be able to describe such statistics.

An open question is whether these results depend upon bulk plasma parameters such as $\beta_{i}$ and, in particular on the angle of $\bold{k}$ to the background magnetic field. A limitation of the current work is that it is based on a handful of data sets with very similar plasma parameters. A significant and maybe conclusive advance would be made if a more comprehensive survey was conducted. This would undoubtedly be a fruitful avenue for future investigations.

\begin{acknowledgments}
	The authors acknowledge the ACE Science Centre; Wind 3DP and MFI teams; Cluster instrument teams for FGM, STAFF-SC, CIS, WHISPER and PEACE; and the DSP group at Rice University for use of the Rice Wavelet MATLAB Toolbox. K. K. acknowledges T. Dudok De Witt, P. Gary and G. Howes for useful discussions and advice on the techniques and physics in this article. O. F. thanks the Comms. \& Signal Processing group in the Dept. of Electrical and Electronic Engineering, Imperial College London, for hosting him as a visiting student. K. K., S. C. and F. S are members of ISSI team 185 `Dissipation/Dispersion in Plasma Turbulence', in the context of which some of the work here was conducted. This work was supported by the UK STFC and the EPSRC. 
\end{acknowledgments}

\appendix

\section{Orthonormal Scale-Dependent Basis and The Undecimated Discrete Wavelet Transform} \label{sec:UDWT} 
We use the Undecimated Discrete Wavelet Transform (UDWT) method (also known as the stationary, translation-invariant, redundant or \emph{a'trous} wavelet transform \citep{Nason1995,Ogden1997,Mallat2009}) to decompose the vector magnetic field observations into scale dependent background and fluctuating components. Unlike the standard discrete wavelet transform (DWT), the UDWT does not downsample the data at each stage of the transform. Thus, the UDWT provides information about the signal at each observation time and so retains event information \citep{Walden1998,PercivalWalden2000}. This timing information is necessary to project the fluctuations onto a mean background field at each observation time. However, the extra redundancy due to no downsampling means that the orthogonality properties of the DWT are no longer present in the UDWT. 

In this article we use the Coiflet 2 wavelet (\emph{coif2}), which provides a $12$-tap high/low pass filter pair, chosen due to the compromise between time-frequency compactness, smoother spectral index estimate, and as its first four moments are zero it captures spectral indices as steep as -9 \citep{Farge2006}. This latter property is very important as the standard method of using increments (equivalent to using a Daubechies order 1 (\emph{db1}), or Harr, wavelet with only one zero moment) to calculate two-point structure functions in turbulence is limited to spectral slopes shallower than -3 \citep{Farge2006,Cho2009}. This is adequate for studying the inertial range of turbulence where the spectral slope is ubiquitously less than -2 \citep{Salem2009}, but is inadequate in the dissipation range where spectral slopes have been observed \citep{Leamon1998,Smith2006} and theoretically predicted \citep{Schekochihin2009} to have much steeper slopes between -3 and -5. Crucially, for data analysis purposes, the filters corresponding to the \emph{coif2} wavelet are virtually symmetric and thus possess a virtually linear phase response \citep{Daubechies1992}. The \emph{coif2} wavelet (and corresponding scaling function) filters used here are identical to the ones published in \citet{Daubechies1992}. 

The UDWT scheme presented below is similar to the wavelet technique pioneered by \citet{Horbury2008} who used a continuous wavelet transform (CWT), where the background magnetic field was defined as convolutions with the scale-dependent envelope of the Morlet wavelet used. However, the method described here differs from the CWT scheme used by \citet{Horbury2008}, as the DWT and UDWT filters used are explicitly designed self-consistently to provide both fluctuations and background fields such that no information is lost and the signal can be reconstructed \emph{exactly} -- a so-called \emph{quadrature mirror filter} \citep{Daubechies1992}. Thus, these filters also ensure that there is no spectral gap between the background field and fluctuations. Details of the actual fast pyramidal algorithms used to implement the UDWT can be found in \citep{PercivalWalden2000} and \citep{Mallat2009}. For this article we simply attempt to summarise what are the salient outcomes of the technique. Thus, the description will be necessarily brief with limited details. A detailed description of the technique as applied to plasma turbulence anisotropy, as well as an extension of this to non-dyadic wavelet packets, will be described in a longer manuscript.

\begin{figure}
	\begin{centering}
		\includegraphics[width=1\columnwidth]{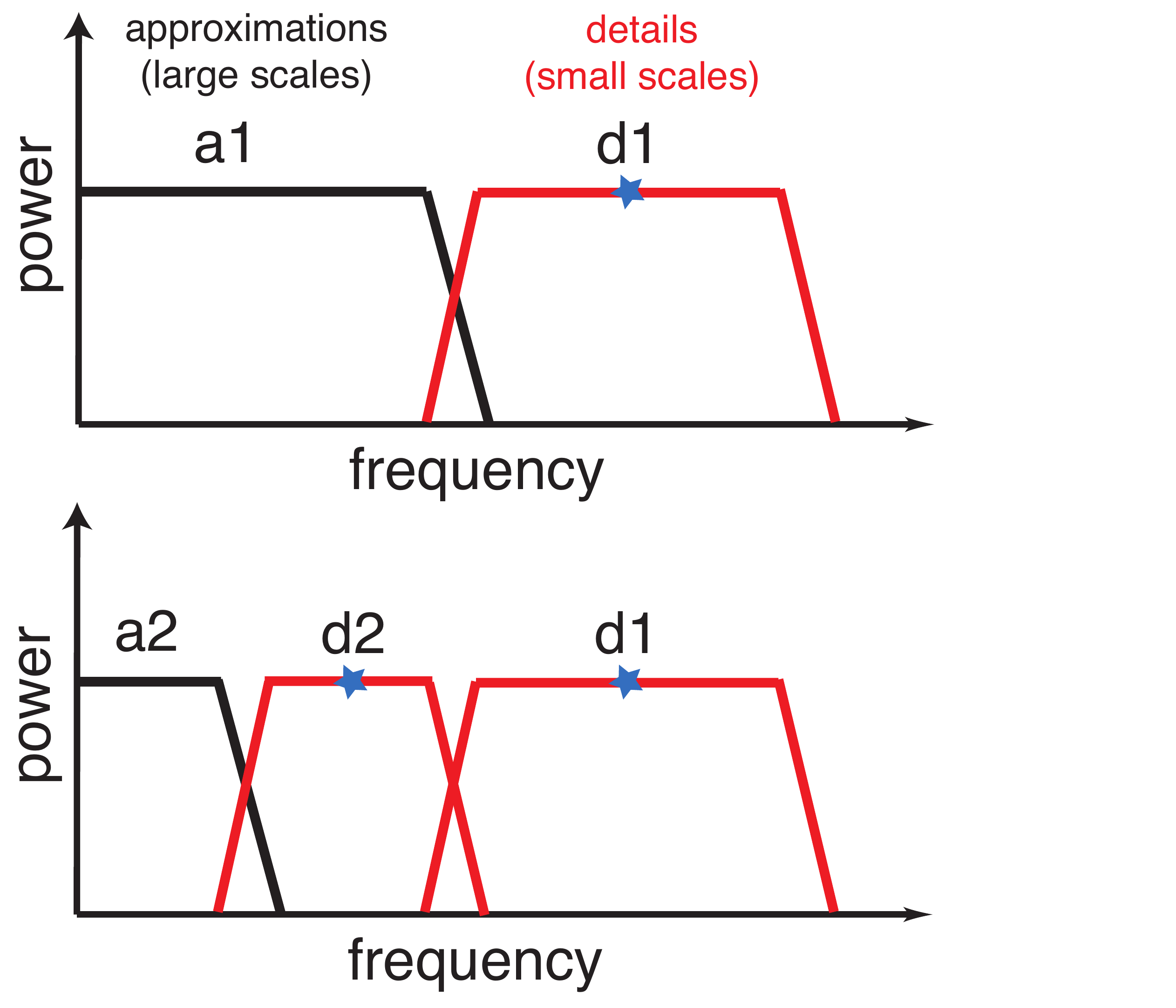} \par
	\end{centering}
	\caption{\label{fig:UDWT}(Upper panel) A schematic, in terms of spectral frames, showing the first stage of a UDWT decomposition in frequency space: the raw signal is decomposed into a low-pass (approximation), `a1', and a high-pass (detail) component, `d1', using the \emph{coif2} wavelet filters (low-high pass quadrature mirror filter pair). Each of these components has an equal (half) share of the frequency space (Lower panel) At the next stage, the `a1' low-pass filtered signal is then further decomposed, into a low-pass signal, `a2', and a high pass-signal, `d2' (again equal share of frequency space). In this way the signal is decomposed into dyadic frequency bands -- equivalent to the application of a dyadic filter bank of band-pass filters to the signal. The resultant set of `details' \{d1,d2,d3 ...\} provide the fluctuations, and the set of `approximations' \{a1,a2,a3 ...\} (once inverse transformed) provide the respective background fields to be projected upon.  The stars represent the central frequency of the resultant band-pass filtered signals used for the fluctuations.}
\end{figure}

The UDWT is equivalent to the following operation: at each temporal scale $\tau$ (corresponding to a frequency) we successively apply a pair of high and low-pass wavelet filters on the data: these provide the `details' (fluctuations) and the `approximations' (precursor to background field). Since these are dyadic filters, they divide the available frequency space in half at each stage. These filters are then upscaled by two and the process is then repeated on the previous approximation signal. The results of the first two stages of this process are shown in fig.\ref{fig:UDWT}. Importantly, at each stage of the UDWT, it is the approximations which are being decomposed into a further set of approximations and details. The effective end result of this successive filtering is a series of \emph{dyadic} band-pass filtered signals of the magnetic field $\mathbf{B}(t_{j})$ (sampled at discrete times $t_{j}$) to give a fluctuation in terms of the wavelet coefficient $\delta\mathbf{B}(t_{j},\tau)$ (at time $t_{j}$ based on a temporal scale $\tau$):
\begin{equation}
	\delta\mathbf{B}(t_{j},\tau)=\sum_{k=1}^{N}\mathbf{B}(t_{k})\left[\sqrt{\tau}\psi(\tau t_{k}-t_{j})\right]\ ,\label{eq:0} 
\end{equation}
where $\tau=2^{i}\Delta:\ i=\left\{ 0,1,2,3,\ldots\right\} $ is the dyadic scale parameter, $\Delta$ is the sampling period of the observations, and $\psi$ is the effective band-pass filter comprised of a succession of the low-pass and high-pass filters. Note that without the factor $\sqrt{\tau}$ and choosing the \emph{db1} (Harr) wavelet instead of the \emph{coif2}, the definition in eq. (\ref{eq:0}) is identical to calculating increments which are used in structure function analysis of intermittency studies in turbulence \citep{Katul2001,Farge2006,Salem2009}. Note also that the resultant wavelet coefficients $\delta\mathbf{B}(t_{j},\tau)$ have to be phase-corrected due to the linear phase shift introduced by the filtering \citep{Walden1998}. Also, edge artifacts due to incomplete convolutions of the signal with the filters, are discarded before the construction of the `approximation' and `detail' signals.

The scale $\tau$ can be related to a central frequency $f$ in Hz \citep{Abry1997} of a dyadic frequency band and so we can write $\delta\mathbf{B}(t_{j},f)$. The result of the low-pass filter at each stage can similarly be written as $\tilde{\mathbf{B}}(t_{j},f)$ using the wavelet conjugate scaling function \citep{Ogden1997}. It is important to note that both $\delta\mathbf{B}(t_{j},f)$ and $\tilde{\mathbf{B}}(t_{j},f)$ live in a function space spanned by the bases constructed from shifts and dilations of the \emph{coif2} wavelet and scaling functions (i.e. shift $t_{j}$ and dilation $\tau$ as in eq. (\ref{eq:0})). We use the inverse UDWT operation on $\tilde{\mathbf{B}}(t_{j},f)$ in order to obtain the mean background field $\overline{\mathbf{B}}(t_{j},f)$, in the direction of which we then project the fluctuations $\delta\mathbf{B}(t_{j},f)$ . 

The mean field direction unit vector is then defined as $\mathbf{e}_{\mathbf{\parallel}}(t_{j},f)= \overline{\mathbf{B}}(t_{j},f)/\left|\overline{\mathbf{B}}(t_{j},f)\right|$. This is distinct from the mean field coordinate presented in \citep{BelcherDavis1971,Leamon1998b} as the mean field here is a locally varying \emph{scale-dependent} field consistent with the scale dependent fluctuations. The other two perpendicular axes are $\mathbf{e}_{\perp1}(t_{j},f)=\left(\mathbf{e}_{\parallel}(t_{j},f)\times\left\langle \mathbf{\hat{V}_{sw}}\right\rangle \right)/\left| \mathbf{e}_{\parallel}(t_{j},f)\times\left\langle \mathbf{\hat{V}_{sw}}\right\rangle \right|$ and $\mathbf{e}_{\perp2}(t_{j},f)=\mathbf{e}_{\parallel}(t_{j},f)\times\mathbf{e}_{\perp1}(t_{j},f)$, where $\left\langle \mathbf{\hat{V}_{sw}}\right\rangle $ is the solar wind velocity direction unit vector time-averaged over the entire interval. Since the solar wind velocity is in a very fast and steady stream, and is within $ $$\sim3^{\circ}$ of the GSE $x$ direction in both the ACE and Cluster intervals, it is reasonable to take a time averaged global, rather than a local, velocity field. Together $[\mathbf{e}_{\perp1}(t_{j},f),\mathbf{e}_{\perp2}(t_{j},f),\mathbf{e}_{\parallel} (t_{j},f)]$ form a time and scale dependent orthonormal basis. The fluctuations parallel $\delta B_{\parallel}(t_{j},f)$ and transverse $[\delta B_{\perp1}(t_{j},f)\ ,\ \delta B_{\perp2}(t_{j},f)]$ to the mean field are then given by projections onto this new basis. 


\bibliographystyle{apj} 

\end{document}